\documentclass{aastex}
\usepackage{emulateapj5,epsfig,graphicx,ifthen}
\newcommand{\figtype}{EPS}
\def\myputfigure#1#2#3#4#5%
{\vskip#5pt\makebox[0pt]{\hskip#2in
\includegraphics[width=#3\textwidth]{#1}}\vskip#4pt\hfill}

\newlength{\colwidth}

\def \sig8 {\hbox{$\sigma_8$} }
\def\spose#1{\hbox to 0pt{#1\hss}}
\def\lta{\mathrel{\spose{\lower 3pt\hbox{$\mathchar"218$}}
     \raise 2.0pt\hbox{$\mathchar"13C$}}}
\def \xray {\hbox{X--ray} }
\def \etal      {et al.\ }
\def \mpc       {{\rm\ Mpc}}
\def \kmsmpc    {{\rm\ km\ s^{-1}\ Mpc^{-1}}}
\def \ergs      { \hbox{$\,$ erg s$^{-1}$} }
\newenvironment{inlinefigure}{
\def\@captype{figure}
\noindent\begin{minipage}{0.999\linewidth}\begin{center}}
{\end{center}\end{minipage}\smallskip}

\shortauthors{BIALEK, EVRARD \& MOHR}
\shorttitle{Simulated Cold Front}

\begin{document}

%\submitted{} \lefthead{BIALEK, MOHR \& EVRARD}
%\righthead{Simulated Cold Front}

\title{ A Cold Front in a Preheated Galaxy Cluster }
\author{ John J. Bialek\altaffilmark{1},  August E. Evrard\altaffilmark{1,2} \& Joseph
J. Mohr\altaffilmark{3}}

\altaffiltext{1}{Department of Physics, 1049 Randall Lab,
University of Michigan, Ann Arbor, MI 48109} 
\altaffiltext{2}{Department of Astronomy, Dennison Building,
University of Michigan, Ann Arbor, MI 48109} 
\altaffiltext{3}{Departments of Astronomy and Physics,
University of Illinois, 1002 W. Green St, Urbana, IL 61801}

\email{jbialek@umich.edu}
\email{jmohr@astro.uiuc.edu}
\email{evrard@umich.edu}

\begin{abstract}

We present a simulated cluster of galaxies, modeled with a pre-heated
intracluster medium, that exhibits X-ray
features similar to the `cold fronts' seen in {\sl Chandra\/}
observations.  Mock observations at a particular epoch
show factor two discontinuities in X-ray temperature and factor
four in surface brightness on a spatial scale $\lta\! 100$~kpc.  
Analysis of the cluster's dynamical history reveals that the
front is a transient contact discontinuity created by an ongoing
merger of two roughly equal mass subgroups.  The cold front feature in
this realization is amplified by the adiabatic expansion of 
one of the subgroups following its ablation from the center of its local 
dark matter potential.  The presence of cold front features in a
cluster modeled without radiative cooling or magnetic fields implies
that such relatively complex physics is not a necessary element of the 
phenomenon and suggests that the prevalence of such features in high
resolution \xray images of clusters may simply reflect the
high frequency of ongoing mergers driven by gravity and comparatively
simple hydrodynamics.  

\end{abstract}


\section{Introduction}

The improved spatial imaging of the hot intracluster medium (ICM) in 
galaxy clusters by the {\sl Chandra\/} Observatory has resulted in the
discovery of apparent contact discontinuities, termed ``cold fronts'', in
many clusters (Markevitch \etal 2000;  Vikhlinin, Markevitch
\& Murray 2001; Mazzotta \etal 2001;  Sun \etal 2002).
These cold fronts occur at the boundary of a local peak in the \xray surface
brightness and are typically located within a few hundred kpc of
the core of the cluster.  The 
fronts exhibit a drop in temperature and corresponding rise in
density upon entering the emission peak.  The term ``cold
front'' is applied to contrast the phenomenon with a
shock front where the temperature would be elevated in the
direction of increasing density.  

The origin of these cold fronts is thought to be related to cluster
mergers (Markevitch \etal 2000).  
From an extended Press--Schechter treatment, Fujita \etal (2002)
predict that up to one-third of present clusters will contain large
\xray subhalos.  
In this {\sl letter}, we present evidence supporting the merger
origin hypothesis from a simulated cluster evolved under a `preheated'
assumption for ICM evolution (Kaiser 1991; Evrard \& Henry 1991).
In this treatment, the proto-ICM gas at high redshift is assumed to
lie a fixed, elevated adiabat that results from heating due to star
formation and/or AGN activity (Bower 1997;  Cavaliere, Menci \& Tozzi
1998; Balogh, Babul \& Patton 1999; Wu, Fabian \& Nulsen 2000; Tozzi
\& Norman 2001; Voit \& Bryan 2001).  The gas subsequently evolves
under gravitationally-driven shock-heating, with 
magnetic fields and radiative cooling ignored.  

As we prepared this work for publication, 
Nagai \& Kratsov (2002), using completely independent techniques,
present cold front phenomena in a pair of simulations of non-preheated
cluster models.  Their simulations have higher spatial resolution
compared to the 
one we present, but our Lagrangian simulations have an advantage in 
allowing the history of gas parcels to be tracked over time.  We use
this ability to show that material once at the core of a merger
progenitor is directly responsible for the cold emission seen at later 
stages of the merger.  

%overview of paper

In section~\ref{sect:sim}, we present the simulation and
compare its cold front properties to observations.  
The merger history and thermodynamics responsible for the cold front
are examined in section~\ref{ssect:phys}.  
Note that all scales quoted in this paper assume a Hubble constant
H$_0 \!=\! 70 \kmsmpc$.  

\section{The Simulated Cluster}
\label{sect:sim}

Following a study of the effects of preheating on cluster scaling
relations (Bialek, Evrard \& Mohr 2001), we have created a new
ensemble of 68 simulated clusters that span roughly a decade in final
mass.  The preheating level of 105.9 keV cm$^2$, favored by the
above study, was imposed upon the ensemble.  All of the
simulations are evolved with the Lagrangian code P3MSPH (Evrard
1988), using the multi-step procedure detailed in Bialek
\etal (2001).  This ensemble has improved resolution relative to
that study; the final hydrodynamic simulation is run on an
effective $96^3$ particle grid.   For the cluster considered in this
letter, the gas and dark matter are represented by
particles of mass $4.43 \times 10^8$ and $3.99 \times 10^9
M_\odot$, respectively.  
The minimum SPH smoothing length is 40 kpc.  Twenty outputs, equally
spaced in proper time, are stored for each cluster.   

    The clusters form in a cold dark matter cosmology
    dominated by vacuum energy density, ${\Lambda}$CDM 
(Efstathiou \etal 2002).  The model assumes a flat spatial 
    geometry with the following parameters: $\Omega_{m} = 0.3$,
    $\Omega_{\Lambda} = 0.7$, $\Omega_{b} = 0.03$, $\sig8 = 1.0$,
    $\Gamma = 0.21$ and $h = 0.7$.  The Hubble constant is defined as
    $100~h \kmsmpc$; and \sig8 is the power spectrum normalization on
    $8h^{-1}$ Mpc scales.

\begin{inlinefigure}
  \ifthenelse{\equal{\figtype}{EPS}}{
   \vbox {
    \parbox{.45\textwidth}{\vskip0.40in
    \centerline{\epsfysize=.9\colwidth \epsfxsize=.9\colwidth\epsffile{figure1a.epsi}}
    }

    \parbox{.45\textwidth}{\vskip0.45in
    \centerline{\epsfysize=.9\colwidth\epsfxsize=.9\colwidth\epsffile{figure1b.epsi}}
   }
   \vfil}
  }
  {\myputfigure{sbin.pdf}{1.8}{1.8}{0}{-55}
    \myputfigure{contourt_rv.pdf}{1.8}{1.8}{-290}{-345}
  }
\figcaption{(a) Grayscale \xray surface brightness map (ranging three
orders of magnitude) with the
general binning structure overlaid (b) Grayscale temperature map
($1 - 3 \times 10^7$ K) with Subgroup~1
contributing surface brightness contours overlaid.
}
\label{plot:contour}
\end{inlinefigure}

\subsection{A Cold Front Feature}
\label{ssect:prop}
%compare a339s observables to Markevitch observations

Examination of some of the mock \xray surface brightness and
temperature maps from the simulation output led to the discovery of a
particular cluster with a strong cold front morphology.  
	This cluster, examined at an
	output epoch $z = 0.22$, has a
	mass of $2.86 \times 10^{14} M_\odot$ resolved by roughly
35,000 gas and 65,000 dark matter particles, \xray
	luminosity of $4.25 \times 10^{43} \ergs$ and an emission weighted
	temperature of 2.3 keV.  All values are quoted 
	within the characteristic radius $r_{200} =1.26 \mpc$
within which the mean interior density is 200 times the critical
density at the redshift of examination.  The different numbers of gas
and dark matter particles reflects the loss of gas within the virial
radius caused by preheating (see Fig.~13 of Bialek \etal 2001).

	Figure~1 shows \xray surface brightness
and temperature maps of the simulated cluster at $z = 0.22$.  The
morphology of the bright \xray emission is highly
distorted, with a horizontal tongue ending in an 
off-center surface brightness peak to the right in the figure.  
The peak in emission is
coincident with a cold spot in the temperature map, a signature feature 
of observed cold fronts. 

Following analysis of Abell 2142 by Markevitch \etal (2000),
 	we construct radial profiles within a restricted arc oriented
along the steep gradient in the surface 
	brightness (to the right of the cold lump, see
	Figure~1a).  The inner two bins are circular,
centered on the	surface brightness peak, and extend to 114 kpc.  The
outer six bins are circular arcs in a wedge with outer radius 460 kpc
	that projects 45 degrees to either side of a horizontal line 
	(parallel to the extended tongue of emission) 
	directed outward from the surface brightness peak.  This geometry is
	used to bin the pixel information in the simulated maps.
	The resulting temperature profile is shown in
	Figure~2a.  
	The cool emission peak extends to approximately 230 kpc; gas
	is a factor 1.5 -- 1.9 cooler  
	inside of the front.  

Both the scale of the cold front
	and the magnitude of the temperature drop are
	consistent with those reported by many observers.
	Observed temperature drops range from about a factor 1.5 seen in
	A1795 (Markevitch, Vikhlinin \& Mazzotta 2001) to about a
	factor of 3 seen in MS 1455.0+2232 (Mazzotta \etal 2001a).
	Both A2142 (Markevitch \etal 2000) and A3667 (Vikhlinin, Markevitch
	\& Murray 2001) are reported to show a factor 2 drop
	in temperature.  Regarding spatial extent, smaller cold fronts
are seen, such as MS
	1455.0+2232 and A2142 which are both about 70 kpc from
	the surface brightness peak.  Likewise, cold fronts as
	large as 271 kpc (RX J1720.1+2638, Mazzotta \etal 2001b)
	and 293 kpc (A3667) are seen.  
Cold fronts resulting from merger activity 
would have sizes determined primarily by the masses of the progenitors.
From the spectrum of progenitor masses expected from hierarchical
clustering (Bower 1991; Lacey \& Cole 1994), one would naturally
expect a range of cold front scales to emerge.   

%density profile
For the set of simulation particles identified in the projected
wedge used to examine the temperature
profile, we derive densities from the normalized emission measure 
$\int \, dV \rho^2/\int \, dV \rho$.  Figure~2b shows
a steep drop in density between 200 and 250 kpc, the region displaying 
the factor two rise in temperature.  The width of this feature is
comparable to the size of the hydrodynamic spatial smoothing at the
interior edge of the front (as illustrated by the bar in
Figure~2b).  
The higher resolution models of Nagai \& Kratsov display smaller 
front widths, but their models do not yet resolve scales $\lta 10$ kpc, 
the width of the front feature observed by Vikhlinin, Markevitch \&
Murray (2001) in A3667.

\begin{inlinefigure}
   \ifthenelse{\equal{\figtype}{EPS}}{
     \centerline{\epsfxsize=.45\colwidth \epsffile{figure2a.epsi}
     \epsfxsize=.45\colwidth \epsffile{figure2b.epsi}}
    }
    {\myputfigure{tempprof.pdf}{-1.4}{0.50}{00}{-10}
      \myputfigure{rhoprof.pdf}{1.8}{0.50}{-35}{-175}
    }
\figcaption{(a) Temperature profile across the cold front.  (b) The
density profile constructed from the emission measure of the particle
distribution.  The
bar in Figure b indicates the average SPH smoothing length in the
region of the cold front.
}
\label{plot:temp}
\end{inlinefigure}

The thermal pressures of particles identified in the projected
wedge show no strong discontinuity across the front.  The velocity
field exhibits 
some radial compression on the far side of the front, but the
associated Mach number is small, $\sim\! 0.1$.  In the next section,
we show that the front is a 
contact discontinuity separating cold, dense material of a subgroup
from the surrounding hotter, more tenuous atmosphere of the combined
cluster.  

\subsection{Cold Front as a Merger Remnant}
\label{ssect:phys}

Examination of outputs earlier than $z = 0.22$ reveals that the
cluster is in the throes of a major merger of nearly equal mass
progenitors.  At a redshift $z = 1.07$, a friends-of-friends group
finder with linking length 0.15 times the mean gas interparticle
separation finds two main subgroups of the cold-front object:
Subgroup~1 has a total mass of $8.0 
\times 10^{13} M_\odot$ and Subgroup~2 has mass $6.4 \times
10^{13} M_\odot$.  The linking length
identifies clumps at densities roughly 500 times the background
value. 

In order to examine the dynamic and thermodynamic histories of gas
within the cores of these subgroups, we employ shorter
linking lengths on the same $z = 1.07$ output.  
Values of 0.06 and 0.10 times the mean
gas interparticle separation used for Subgroups 1 and 2
result in core gas masses of $3.4 \times 10^{11} M_\odot$ and $2.1 \times
10^{11} M_\odot$ respectively. The
Lagrangian nature of the computation allows the core material in these
subgroups to be tracked across all output epochs.  

At $z=0.22$, the subgroup cores identified at $z=1.07$
continue to define coherent gas clumps embedded within the larger
atmosphere of the merging cluster.  Their centroids lie at opposite ends of the
tongue of emission shown in Figure~1a.  
In Figure~1b, we superpose the \xray emission 
from the core gas associated with Subgroup~1.  This material 
is responsible for the surface brightness peak and temperature dip at
the location of the cold front.

Examination of the orbit of the subgroups shows that they 
orbit each other with significant angular momentum oriented nearly
vertically in the plane of the images shown in Figure~1.  The subgroup
orbits projected along the rotational axis direction are shown in 
Figure~3.  The location of the most bound gas
particle associated with each subgroup is plotted for
outputs before and after the viewing redshift $z=0.22$ (indicated by solid
symbols).  The cores of the subgroups avoid direct collision and the
cold front occurs at apogee of their orbit.

\begin{inlinefigure}
   \ifthenelse{\equal{\figtype}{EPS}}{
     \epsfxsize=.45\textwidth
     \epsfysize=.45\textwidth
     \centerline{\epsfysize=\colwidth\epsfxsize=\colwidth\epsffile{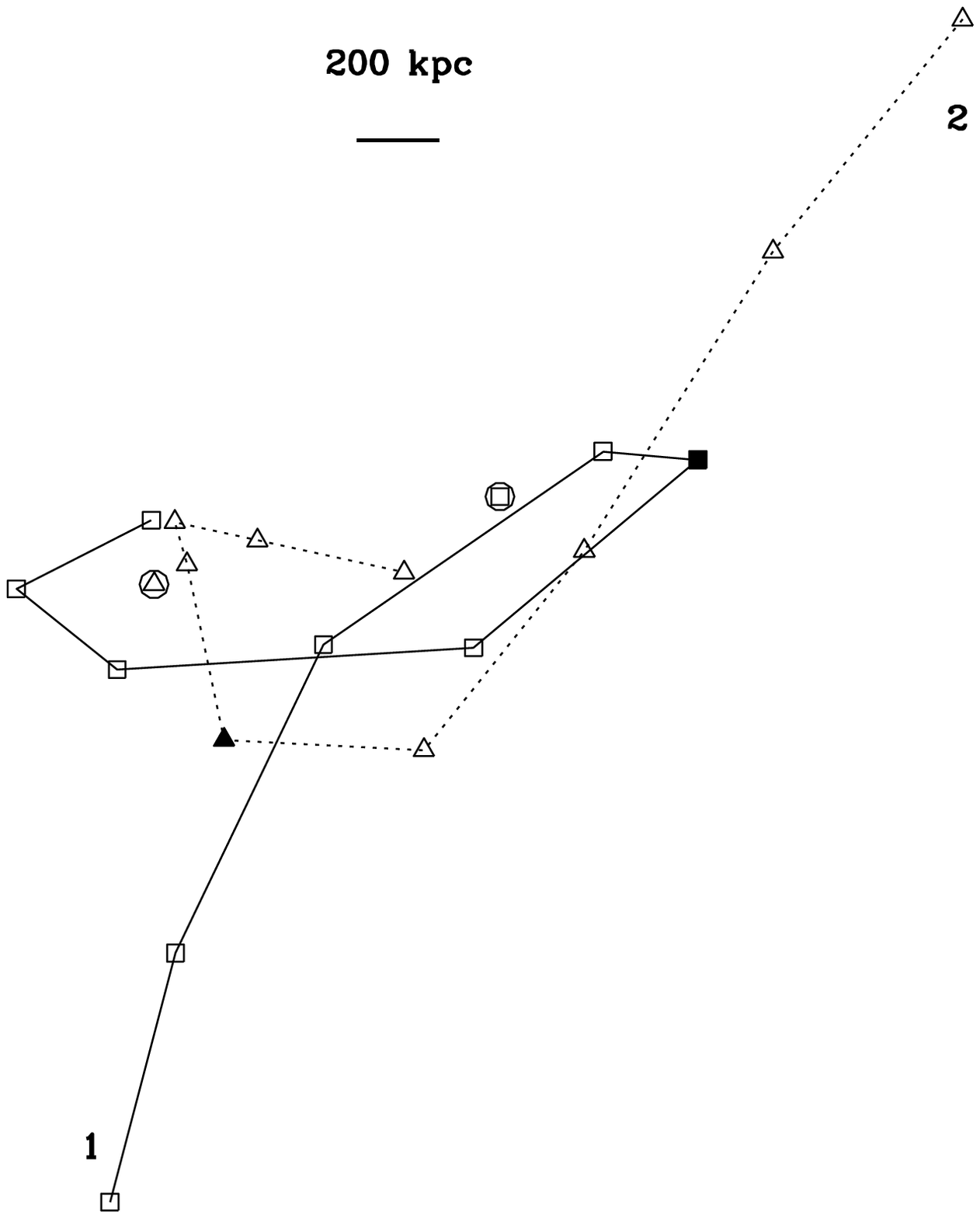}}
  }
  {\myputfigure{posmb.pdf}{0.0}{1.0}{-70}{-30}
  }
\figcaption{Orbits of the subgroup cores in comoving coordinates.  
The most bound gas particle
associated with each subgroup is plotted over 9 outputs ranging from
$z=0.54$ to 0.   
The earliest epoch is labeled by subgroup number and 
the $z=0.22$ output, in
which the cold front is examined, is indicated by solid symbols.
The symbols enclosed by circles indicate the location of the most
bound dark matter particle at the viewing epoch.  The solid
circle indicates the location of the most bound particle for the
overall cluster at the viewing epoch.  
The bar on the top right indicates a 200 kpc scale.  The projection used in
Figure~1 is for an observer located in the plane of
the orbit and viewing the merger from below.
}
\label{plot:orbit}
\end{inlinefigure}

During the encounter, the gas of Subgroup~1 is ablated away from the
core of its local confining dark matter potential.  At the viewing
output, Subgroup~1's gas lags behind its collisionless dark matter
component by 300 kpc (see Figure~3).  Having been
deposited into a shallower part of the potential,
the overpressured gas expands adiabatically and cools.  The cold
nature of the core emission arises from this merger-induced
adiabatic expansion. 

Subgroup~2, which at as early as $z=0.54$ is somewhat more diffuse than
Subgroup~1, experiences different dynamics and does not survive the
interaction as comparatively intact.  At the viewing output, its gas
component has spread out over about a factor two in length and it 
experiences some increase in entropy due to shock heating.  

The different thermodynamic behavior of the two subgroups is shown in 
Figure~4, where we plot logarithmic changes in density 
and temperature experienced by the subgroup gas over a time period of 
538 Myr (the output time interval of the computation) preceding the
viewing epoch $z=0.22$.  Just prior to the viewing epoch, particles in 
Subgroup~1 experience a drop in density and consequentially cool,
sliding down the $\gamma = 5/3$ adiabat shown in the figure as the
straight line.  The gas of Subgroup~2, which has also
been left behind by its dark matter, is in a different state.  
The leading edge of the gas cloud is becoming more dense as it falls
into its dark 
matter potential well and is gaining entropy as it is being
mildly shock heated.  This leading edge is leaving behind a diffuse tail 
of lagging gas, some of which is expanding and cooling adiabatically
like the  gas of Subgroup~1.

\begin{inlinefigure}
\medskip
   \ifthenelse{\equal{\figtype}{EPS}}{
     \epsfxsize=.45\textwidth
     \epsfysize=.45\textwidth
      \centerline{\epsfysize=\colwidth\epsfxsize=\colwidth\epsffile{figure4.epsi}}
   }
   {\myputfigure{delphase.pdf}{0}{1.0}{-70}{-30}
   }
\figcaption{Logarithmic changes in density and emperature for
particles originating in the cores of Subgroups~1 and 2 (labeled
accordingly) over a 538 Myr interval preceding the $z=0.22$ viewing
epoch.  The superimposed lines marks zero change in entropy.  Note
that the Subgroup~2 data are offset by 0.5 for clarity.
}
\label{plot:dphase}
\end{inlinefigure}

From analysis of the pair of outputs that precede the viewing epoch, we
find that most of the core gas in Subgroup~2 does undergo a period of
adiabatic expansion similar to that of Subgroup~1.  In the output
preceding $z=0.22$, a pair of cold spots appear in the
temperature map, but their amplitudes, as well as the amplitudes
of their corresponding emission peaks, are somewhat smaller than 
the features displayed at the viewing epoch of Figure~1.  The
adiabatic expansion experienced by the core of Subgroup~2 also occurs
during the ablation of the gas away from the center of its local
gravitational potential.  Spatial displacement of cold front gas from 
the local dark matter peak is also noted in the simulations of Nagai
\& Kratsovv (2002).  Adiabatic cooling of ablated dense material,
coupled with compression and heating of less dense cluster gas in
forward of the ablated core, may be the key ingredients involved in
the creation of cold fronts by mergers.  

Sun \etal (2002) provide observational support for the separation of 
core gas from its local gravitational potential.  
They find a scarcity of galaxies within the observed cold front
and suggest that the gas is lagging behind its associated
galaxies.  Assuming that galaxies follow trajectories like those
defined by the collisionless dark matter, this conjecture is in
agreement with the separation of  
the gas and dark matter seen in the simulation. 

\section{Conclusion}

We show that clusters modeled under the assumption of a preheated
intracluster medium can exhibit features similar to the cold fronts
observed in high resolution spectroscopic imaging of \xray emission
from clusters.   We present a particular realization, displaying a
temperature dip coincident with a peak in \xray
surface brightness, where the cold front is a transient feature
created by ablated core material of a merging subgroup.   Freed from its
confining local dark matter potential, adiabatic expansion cools the
core while its density remains sufficiently high to create a strong
feature in emission.  This result supports the merger origin
assumption for cold fronts and demonstrates that their observed 
characteristics can be reproduced by a gas dynamic treatment that
ignores radiative cooling and magnetohydrodynamics.  A similar
conclusion using independent methods is reached by Nagai \& Kratsov
(2002). 

Although these studies provide an existence proof that 
cold fronts can result from mergers, we do not yet know if {\sl all}
observed cold fronts are consistent with this formation mechanism.  
It remains to be seen what fraction of mergers result in cold 
fronts and what combination of parameters --- mass ratio, impact
parameter, angular momentum, viewing angle --- favors such an outcome.
Although the cold front in this study occurs when the
subcluster's gas strays from its local potential minimum and expands
adiabatically, we do not know if this is a necessary condition.  

Improved understanding of cold fronts will require more extensive 
searches within well-defined samples of observed and simulated clusters.  
The relatively small numbers of observed and simulated cold fronts
must be increased to enable secure statistical studies of this
phenomenon in the cluster population.  

\acknowledgments

This simulation was produced using the computing facilities
at the University of Illinois's National Center for
Supercomputing Applications.   This work was supported by
NASA through grant NAG5-7108 and NSF through grant AST-9803199.

\end{document}